\documentclass[12pt,draftclsnofoot,onecolumn,twoside]{IEEEtran}
\usepackage[mathscr]{eucal}
\usepackage{ifpdf}
\usepackage{cite}
\usepackage{graphicx}
\usepackage[cmex10]{amsmath}
\usepackage{amssymb}
\usepackage{algorithmic}
\usepackage{units}
\usepackage{setspace}
\usepackage{algorithm}
\usepackage{array}
\usepackage[tight,footnotesize]{subfigure}
\usepackage{amsthm}
\usepackage{slashbox}
\usepackage{multirow}
\usepackage{enumerate}
\usepackage{color}
\newcounter{assume}

\newtheorem{theorem}{Theorem}
\newtheorem{lemma}{Lemma}
\newtheorem{proposition}{Proposition}

\newtheorem{remark}{Remark}

\newtheorem{definition}{Definition}

\begin{document}
\title{Distributed User Association in Energy Harvesting Small Cell Networks: A Probabilistic Model}
%
\author{
\IEEEauthorblockN{Setareh Maghsudi and Ekram Hossain, \textit{Fellow, IEEE}\\}
\thanks{The authors are with the Department of Electrical and Computer Engineering, University 
of Manitoba, Winnipeg, MB, Canada (e-mails: \{setareh.maghsudi, ekram.hossain\}@umanitoba.ca).}
}
\maketitle
%
\begin{abstract}
We consider a distributed downlink user association problem in a small cell network, where small 
cells obtain the required energy for providing wireless services to users through ambient energy 
harvesting. Since energy harvesting is opportunistic in nature, the amount of harvested energy is 
a random variable, without any a priori known characteristics. Moreover, since users arrive in the network 
randomly and require different wireless services, the energy consumption is a random variable as 
well. In this paper, we propose a probabilistic framework to mathematically model and analyze the 
random behavior of energy harvesting and energy consumption in dense small cell networks. Furthermore, as 
acquiring (even statistical) channel and network knowledge is very costly in a distributed dense 
network, we develop a bandit-theoretical formulation for distributed user association when no 
information is available at users. 
\end{abstract}
%
\section{Introduction}
\label{sec:Introduction}
In order to cope with the ever-increasing need for mobile services, future wireless networks are foreseen to deploy dense small cells to underlay the legacy macro cellular networks. This concept takes advantage from low power short-range base stations that offload macro cell traffic \cite{Hossain14:ETG}, \cite{Fodor12:DAN}. As usual, these advantages come at some cost; more specifically, system designers face a variety of new challenges in order to realize the concept of small cell networks. Examples of challenges include synchronization \cite{Zou15:NSDS}, resource allocation \cite{Amin15:APG}, interference mitigation \cite{Zhang15:ROIM}, handover management \cite{Zhang15:CIM}, and user association, which is the focus of this paper, as described in the following.
\subsection{Motivation and Contribution}
\label{subsec:Contribution}
User association is a fundamental problem in wireless communications that has been under intensive investigation in the past decade; however, due to structural differences between dense small cell networks and conventional cellular networks, the association methods developed to be applied in the latter might not be efficient when used in the former; consequently, it becomes imperative to search for new approaches that are specifically tailored for the emerging networking concepts, including 5G small cell networks. In the following, we review important existing works. In 
\cite{Semiari14:MTP}, matching theory is applied to solve the user association problem in dense small cell networks. A similar work is \cite{Namvar14:ACM}, where the authors propose a context-aware user-cell association approach that exploits the information about the velocity and trajectory of users. While taking the quality of service (QoS) requirements into account, matching theory is used to design a novel algorithm to solve the user association problem. Reference \cite{Saad14:CAG} formulates the uplink user association as a college admission game and proposes an algorithm based on coalitional games to solve the problem. Joint user association and resource allocation is investigated in 
\cite{Chen15:JUA}, and a belief propagation algorithm is proposed for joint user association, sub-channel allocation, and power control. Energy-efficient and traffic-aware user association are studied in \cite{Mesodiakaki14:EEU} and 
\cite{Elbassiouny15:TAU}, correspondingly. The results show that exploiting the available context-aware information, for example, users' measurements and requirements, as well as knowledge of the network, can improve energy- and spectrum-efficiency when performing the user association. A cross-layer framework for user association control in wireless networks is investigated in \cite{Athanasiou09:ACL}. Load balancing through efficient user association is investigated in 
\cite{Ye13:UAL}. In a large body of previous research works, the proposed user association scheme is centralized or only partially distributed, which necessitates the availability of global channel state information (CSI) at least at a central node, resulting in high computational cost and/or overhead. Therefore, it is necessary to develop distributed user association schemes that are able to cope with information shortage.

Furthermore, in a dense small cell network, unlike the conventional cellular networks, small cells are irregularly deployed; hence, not all of them can be connected to a power grid. Therefore, the required energy for small cells may need to be harvested locally from the ambient environment \cite{Sudeva11:EHS}, rather than being provided by using a fixed power supply. By using this concept, not only the small cells become self-healing but also frequent recharge of fixed power supply and/or the cost and waste of transferring the energy from a power beacon can be avoided. This sort of energy-independence is in particular feasible in small cell networks, since small cells normally provide limited services to a small number of users; that is, the energy obtained through energy harvesting might suffice to satisfy users' requirements. Nonetheless, since energy harvesting is opportunistic in general, uncertainty is a natural attribute of the amount of residual energy in small cells. In the presence of uncertainty, distributed user association becomes even more challenging, since assignment is performed \textit{before} any information regarding the amount of energy in each small cell is disclosed. 

In a vast majority of existing literature, the proposed user association method is designated for a specific energy harvesting model, for example, random Poisson process \cite{Song14:TUA} or Bernoulli energy arrival \cite{Yu15:EHP}. Nonetheless, according to \cite{Lee11:EMS}, many distributions such as geometric distribution, Poisson distribution, transformed Poisson distribution as well as Markovian model are not adequate to model the random harvested energy; as a result, it is important to look for new analytical models for random energy harvesting, which includes a combination of distribution functions. In addition, it is clear that a strong dependency between the user association method and the model of energy harvesting reduces the method's applicability. User association in conjunction with energy harvesting in small cell network is also considered in \cite{Sakr15:AMT}. Therein, stochastic geometry is used to develop a modeling framework for $K$-tier uplink cellular networks with RF energy harvesting from the concurrent cellular transmissions.

In this paper, we consider a distributed small cell network with energy harvesting, where all network characteristics, including frequency of energy arrival, energy intensity, quality of wireless channels, as well as user arrival at every small base station 
(SBS), are non-deterministic and hence, uncertain. This stands in sharp contrast with most previous works, in which only some of the network characteristics are regarded as random variables. We develop a new analytical model for energy harvesting, and we define the notion of successful transmission under uncertainty. We then derive a formula for success probability in this random environment. Assuming that no central controller exists and also users are not provided with any channel and network information, we cast the distributed user association problem as a multi-armed bandit problem with sleeping arms, and we solve the formulated problem using some algorithmic solution. Unlike many previous works, the proposed user allocation scheme is distributed, does not require any information at users, and does not depend on the specific model of energy harvesting; thus it is highly flexible and offers more applicability in comparison with state-of-the-art solutions. 
\subsection{Paper Organization}
\label{subsec:Organization}
The paper is organized as follows. In Section \ref{sec:SysMod}, we describe the small cell network model 
together with energy harvesting and transmission protocols. In Section \ref{sec:energyModel}, we propose 
two probabilistic models to analyze energy harvesting and energy consumption in small cell networks. In 
Section \ref{sec:Problem} we present the user association problem. Bandit-theoretical model of the formulated 
user association problem is described in Section \ref{sec:Bandit}. Section \ref{sec:Numeric} includes 
numerical analysis and discussions. Section \ref{sec:Conclusion} concludes the paper.
\section{System and Transmission Model}
\label{sec:SysMod}
We consider a dense small cell network consisting of a set $\mathcal{M}$ of $M$ small cells and a set $\mathcal{N}$ of $N$ users. Data packets are transmitted to the users in the downlink in successive  transmission rounds. For every transmission round, each user is associated to only one\footnote{However, as will be discussed later in this paper, the proposed solution is also applicable to the case where every user might associate to multiple SBSs of its choice.}~small cell of its own choice. That is, every user selects an SBS by itself, which implies that the association is performed in a distributed manner. Multiple users can be served by a single SBS. By $\mathcal{N}_{m}$ we denote the set of $N_{m}$ users to be served by SBS $m \in \mathcal{M}$. For the transmission of every data packet, every user $n$ requires a specific quality of service (QoS) that is expressed in terms of a minimum data rate $r_{n,\min}$. If communicating via SBS $m \in \mathcal{M}$, the QoS of a user $n$ is satisfied when it is allocated some energy $q_{nm}$. As mentioned before, unlike conventional cellular infrastructures, in a small cell network, SBSs are irregularly deployed so that many of them cannot be attached to a power grid. Therefore, we investigate a scenario in which every small cell obtains the energy through local ambient energy harvesting, for instance, by attracting and converting the solar or wind energy. We assume that energy harvesting is independent across small cells. Since energy harvesting is random in nature, in each small cell the amount of harvested energy is a random variable. We assume that every SBS uses a \textit{harvest-use} strategy. For an SBS $m \in \mathcal{M}$, this scheme is briefly described in the following. 

The SBS operates \textit{periodically} in two consecutive steps, namely, energy harvesting (inactive) and data transmitting (active). In the first step, which lasts for some time denoted by $T_{m}$, the SBS harvests the energy. During this time, that is, before any information about the amount of harvested energy is disclosed, SBS $m$ is selected by some users, in a distributed manner, as service provider. At the end of energy harvesting step, the SBS announces to the network (for instance, by using a broadcast signal), that it starts the second step, i.e., it enters the active mode. Transmissions are performed in the second step, which lasts until either the energy is exhausted or all assigned users are served. The end of this step and re-entering the inactive mode is also announced to the network. For simplicity, we assume that no energy is stored and transferred from one period to the other. In other words, the number of users to be served by every SBS is large enough so that the residual energy at the end of second step can be neglected compared to the newly harvested energy. Every SBS allocates energy to users on a first-come first-served basis; therefore, the number of users that can be served by every SBS depends on the amount of harvested energy. We assume that at every SBS $m$, the allocated energy to each user cannot exceed a maximum amount, say, $q_{m,\max}$. Intuitively, this assumption improves the energy efficiency of the network by providing incentive to users to select an SBS to whom they have high channel quality, so that the required energy does not exceed the threshold. In case the required energy is larger than the allowed amount, transmission is still performed, but clearly with some quality of service lower than requested. 

We assume that each small cell is provided with sufficient spectrum resources to guarantee orthogonal transmission to its assigned users; that is, inside every small cell, transmissions are corrupted only by zero-mean additive white Gaussian noise (AWGN) with variance $N_{0}$. For each small cell $m \in \mathcal{M}$, the intercell interference experienced by every user $n \in \mathcal{N}_{m}$, denoted by $I_{nm} \geq 0$, is regarded as noise and is assumed to be fixed and known. The real-valued channel coefficient between 
user $n \in \mathcal{N}_{m}$ and small cell $m \in \mathcal{M}$ is denoted by $h_{nm}$. We assume frequency non-selective  block fading channel model, where $h_{nm}$ is Rayleigh-distributed and remains fixed during the transmission of every packet for all $n \in \mathcal{N}$ and $m \in \mathcal{M}$.\footnote{Although we focus on Rayleigh fading model for our analysis of energy consumption, the proposed association method does not depend on the channel fading model.}~For each $n \in \mathcal{N}_{m}$, the achievable transmission rate is given by
\begin{equation}
\label{eq:Rate}
r_{nm}(h_{nm})=\log \left(1+ \frac{P_{nm}\left| h_{nm} \right|^{2}}{N_{0}+I_{nm}} \right),
\end{equation}
where $P_{nm}$ is the transmit power of SBS $n$ to user $m$.

Since in small cell networks the number of SBSs is large, the user cannot acquire the statistical information of 
all channels to all SBSs. Consequently, in order to make the model realistic, we assume that at the time of SBS selection, the user does not have any information about channel quality, amount of harvested energy and/or network traffic. After the energy harvesting (inactive) step and at the beginning of transmission (active) step, every SBS acquires the channel state information (CSI) of assigned users by using pilot signals, in order to allocate the required energy. This task is performed sequentially according to the selection order, i.e., on a first-come first-served basis. The SBS stops as soon as all energy is allocated, and the remaining users are denied services. Transmission is performed either sequentially or simultaneously, depending on the number of antennas and frequency resources available at the SBS. For every SBS $m$, the energy harvesting and transmission protocol is summarized in \textbf{Algorithm \ref{Alg:Trans}}. Note that SBSs and/or users are not required to be synchronized.
\begin{algorithm}
\caption{Energy Harvesting and Transmission Model}
\label{Alg:Trans}
\small
\begin{algorithmic}[1]
\FOR{Period $j=1,2,...$}
\STATE For $0<t<T_{m}$, 
       \begin{itemize}
        \item The SBS harvests energy. 
        \item Given no information, a set $\mathcal{N}_{m}$ of users selects SBS $m \in \mathcal{M}$ for transmission 
             (Section \ref{sec:Bandit}).  
       \end{itemize}
\STATE By using a broadcast signal, the SBS announces to the network that it enters the active (transmission) mode. 			
\STATE For $t>T_{m}$, 
       \begin{itemize}
        \item The SBS knows the amount of its harvested energy. 
        \item On a first-come first-served basis, the SBS serves its assigned users as follows: 
        \begin{itemize}
           \item It obtains CSI by using pilot signals;
           \item By signaling from the user, it acquires the required QoS information; 
           \item It calculates and allocates the required energy;
           \item Transmission is performed. 
       \end{itemize}
       \end{itemize}
\STATE By using a broadcast signal, the SBS announces to the network that it enters the inactive (energy harvesting) mode.			
\ENDFOR       
\end{algorithmic}
\end{algorithm}
\begin{remark}

Traditionally, each user can associate to a single SBS; multiple simultaneous associations, however, would enhance the system throughput and reduce the outage ratio, particularly for cell edge users. In contrast to most previous works, our proposed user association scheme is also applicable to the network model in which every user $n \in \mathcal{N}$ is allowed to associate to multiple SBSs, say, a set $\mathcal{M}_{n} \subseteq \mathcal{M}$ with cardinality $M_{n}$. In particular, in Section \ref{sec:Bandit}, we will describe that by using the proposed selection policy, user $n$ simply selects 
$M_{n}$ SBSs instead of one SBS only. Such user can be thus regarded as multiple (i.e., $M_{n}$) virtual users, each of them associated to a single SBS. Clearly, this interpretation gives rise to invisible changes in network characteristics; for instance, visible network traffic is lighter than the true one, as every physical user that arrives in the network would act as multiple virtual users. Nonetheless, imprecise network characteristics do not affect the performance of the proposed selection method due to the following reason: As we will see later, selections are performed in a distributed manner by users, which are assumed to have no prior information. More precisely, all \textit{true} network characteristics are learned through successive interactions with the environment. Therefore, the hidden effects are learned as well.
\end{remark}
\section{Analytical Models of Energy Harvesting and Energy Consumption}
\label{sec:energyModel}
Before proceeding to the user association problem, in this section we describe the analytical models of energy harvesting as well as energy consumption.
\subsection{Energy Harvesting}
\label{subsec:energyHar}
Intuitively, energy harvesting is of opportunistic nature; as a result, the amount of harvested energy is 
a random variable, which, may not be easily attributed to some well-known probability distribution 
function. In fact, according to \cite{Lee11:EMS}, many distributions such as geometric distribution, Poisson 
distribution, transformed Poisson distribution as well as Markovian model are not adequate to model the random 
harvested energy, and a combination of distribution functions should be used for analytical modeling. In this 
paper, we propose to use a compound Poisson model for energy harvesting, as described in the following. 

For every SBS $m \in \mathcal{M}$, the energy arrival, $K_{m}$, is modeled by a Poisson Process with rate $\lambda_{m}$; that is, $K_{m} \sim \textup{Poi}(\lambda_{m})$. Moreover, at every arrival, the amount of harvested energy, denoted by 
$X_{m,i}$, is modeled as a random variable following exponential distribution with parameter $\mu_{m,i}$; i.e., 
$X_{m,i} \sim \textup{Exp}(\mu_{m,i})$. We assume that each SBS continues to harvest the energy until the 
$k_{m}$-th arrival. If $\mu_{m,i}$ is known at SBS $m$, $k_{m}$ can be selected according to its storage capacity; otherwise it is simply selected randomly. Afterward, transmission (active) step starts. As a result, the duration of the energy harvesting (inactive) step, $T_{m}$, is a random variable itself. As it is well-known, for any Poisson process with rate $\lambda$, the inter-arrival time follows an exponential distribution with parameter $\lambda$. Thus, $T_{m}$ has the distribution of the sum of $k_{m}$ independent and identically-distributed (i.i.d.) exponential random variables, which, according to the following lemma, is an Erlang distribution with parameters $k_{m}$ and $\lambda_{m}$, i.e, $T_{m} \sim \textup{Erl}(k_{m},\lambda_{m})$. 
\begin{lemma}[\cite{Amari97:CFE}]
\label{lm:ExpSumIid}
Let $X_{i}$, $i \in \left \{1,...,k \right \}$, be i.i.d. random variables, where $X_{i} \sim 
\textup{Exp}(\lambda)$. Then $S=\sum_{i=1}^{k}X_{i}$ follows an Erlang distribution with parameters 
$k$ and $\lambda$, i.e., $S \sim \textup{Erl}(k,\lambda)$, so that
\begin{equation}
\label{eq:Erlang}
f_{S}(s)=\frac{\lambda^{k}}{(k-1)!}s^{(k-1)}\textup{exp}(-\lambda s).
\end{equation}
\end{lemma}
Physically, this model can be explained as follows. An arrival corresponds to an event when energy harvesting is possible; the amount of energy, however, is not equal at all arrivals. For instance, assume that the energy is harvested from the wind by using anemometer. When the wind intensity is larger than a specific threshold, then some energy can be obtained. A higher wind intensity, however, results in larger amount of energy, and \textit{vice versa}. We model this phenomenon by using exponential distribution, since in most environments intensive weather conditions are unlikely; that is, at a single event, it is unlikely that the SBS harvests a very large amount of energy. Moreover, this model implies some sort of worst-case analysis, since in any exponential distribution with some fixed parameter, smaller values are more likely to happen than larger ones. The required number of arrivals to fill the storage capacity can be selected based on the weather forecast. It should be noted that, in this paper, the user association scheme does not assume any information on energy harvesting or channel quality, and therefore is not affected by the probabilistic model of energy harvesting. 

Now we are in a position to formalize the proposed energy harvesting model. Let $Y_{m}$ be the stored energy at small cell $m \in \mathcal{M}$, at the end of energy harvesting period. Then we have
\begin{equation}
\label{eq:Energy}
Y_{m}=\sum_{i=1}^{k_{m}}X_{m,i},
\end{equation}
where, by the discussion above, $k_{m}$ is the required number of energy arrival events to stop the inactive mode, 
and $X_{m,i} \sim \textup{Exp}(\mu_{m,i})$ is the amount of harvested energy at the $i$-th event. In what follows, we derive the probability density function of $Y_{m}$. In doing so, we distinguish the following two cases: i) Energy arrivals are independent and identically-distributed; ii) Energy arrivals are independent, but distributions are not identical. 
\subsubsection{Independent, identically-distributed energy arrivals}
\label{ssubsec:i.i.d-HarEnergy}
Let the intensity of energy arrivals be modeled by i.i.d. random variables so that $\mu_{m,i}=\mu_{m}$ for 
$i \in \{1,...,k_{m}\}$. Then by (\ref{eq:Energy}) and according to Lemma \ref{lm:ExpSumIid}, we have $Y_{m} 
\sim \textup{Erl}(k_{m},\mu_{m})$.\\ 
\textbf{Normal Approximation-} According to the central limit theorem, in case $k_{m}$ is large enough, for instance 
$k_{m}>30$, $Y_{m}$ can be approximated by a normal random variable with mean $\frac{k_{m}}{\mu_{m}}$ and variance 
$\frac{k_{m}}{\mu_{m}^{2}}$; i.e., $Y_{m} \sim \textup{Nor}(k_{m}/\mu_{m}, k_{m}/\mu_{m}^{2})$. 
\subsubsection{Independent, non-identically-distributed energy arrivals}
\label{ssubsec:i.ni.d-HarEnergy}
Before proceeding to calculate $f_{Y}(y)$ for independent but non-identical (i.ni.d.) $X_{t}$, 
we state the following lemma. 
\begin{lemma}[\cite{Yao90:OPA}]
\label{lm:ExpSum}
Let $X_{i} \sim \textup{Exp}(\mu_{i})$, $i \in \left \{1,...,k \right\}$, and $S=\sum_{i=1}^{k}
X_{i}$. The probability density function of $S$ is given by
\begin{equation}
\label{eq:ExpSumN}
f_{S}(s)=\sum_{i=1}^{k}A_{i}e^{-\mu_{i}s}, 
\end{equation}
with 
\begin{equation}
A_{i}=\prod_{j=1,j\neq i}^{k}\frac{\mu_{j}}{\mu_{j}-\mu_{i}}. 
\end{equation}
\end{lemma}
Thus, if energy arrivals are not identically-distributed, $f_{Y}(y)$ can be concluded from Lemma \ref{lm:ExpSum}.\\
\textbf{Normal Approximation-} In i.ni.d. case, the central limit theorem can be still applied, provided that the 
\textit{Lyapunov condition} is satisfied \cite{Billingsley86:PaM}. Roughly speaking, the condition implies that for large enough $k_{m}$, the contribution of every $X_{m,i}$ to the sum $Y_{m}$ is limited. Then, 
$Y_{m} \sim \textup{Nor}(\sum_{i=1}^{k_{m}} 1/\mu_{i,m},\sum_{i=1}^{k_{m}} 1/\mu_{i,m}^{2})$.     

\begin{figure}[t]
\centering
\includegraphics[width=0.45\textwidth]{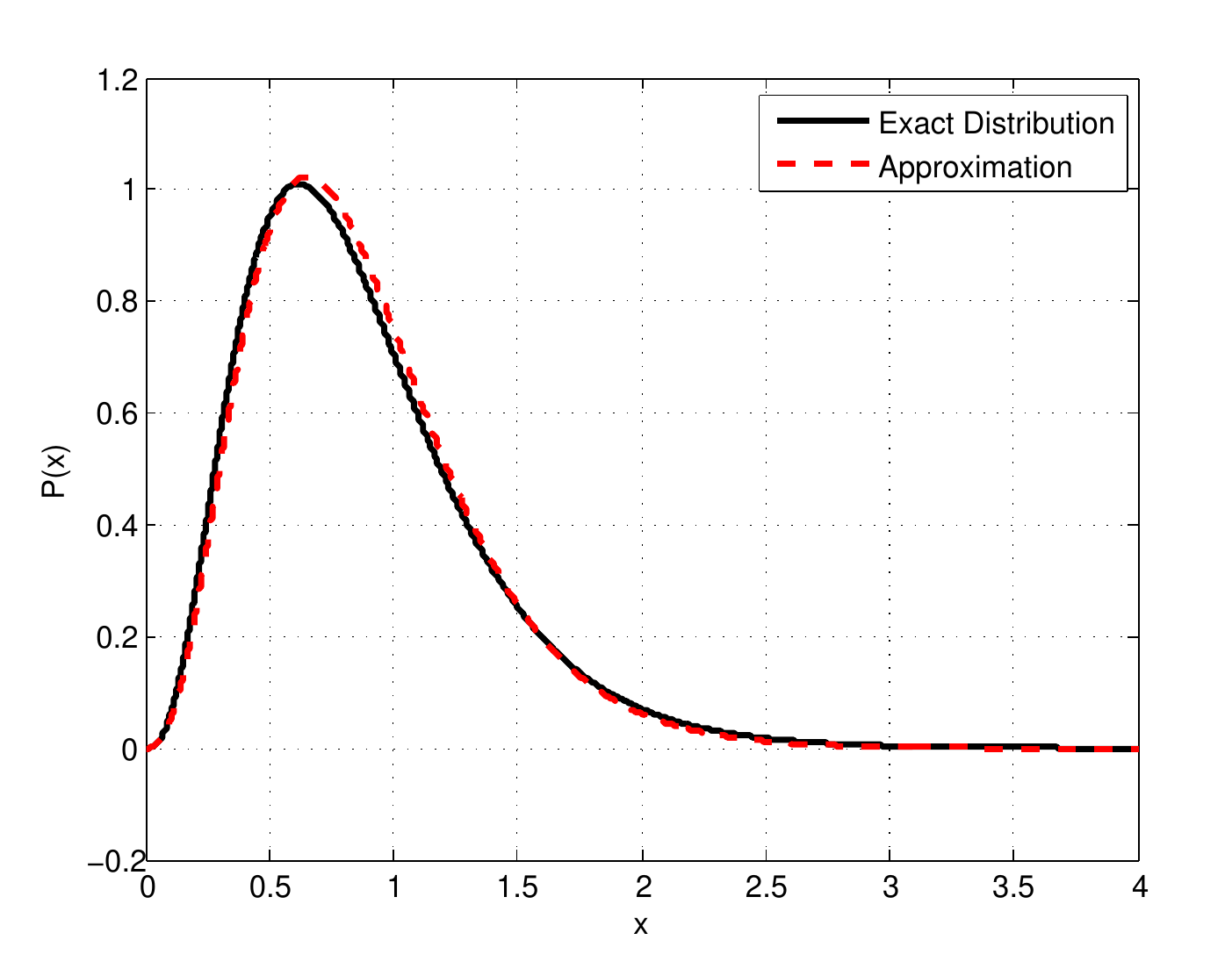}
\caption{Approximation of $f_{S}(s), S=\sum_{i=1}^{4}X_{i}$. $X_{i}$, $i \in \left \{1,...,4\right\}$ are exponential random variables with $\mu_{1}=3$, $\mu_{2}=4$, $\mu_{3}=6$, and $\mu_{4}=8$.}
\label{Approx}
\end{figure}

It is clear that the distribution expressed in (\ref{eq:ExpSumN}) is difficult to trace. In what follows, 
we describe a condition under which i.ni.d. sums can be approximated by i.i.d. sums, so that for the 
i.ni.d. case, $f_{Y}(y)$ can be approximated by an Erlang distribution, given in (\ref{eq:Erlang}). To this end, we proceed to the following proposition.
\begin{proposition}
\label{Pr:Approx}
Let $S=\sum_{i=1}^{k}X_{i}$, where $X_{i} \sim \textup{Exp}(\mu_{i})$. Also, let $Q=\sum_{i=1}^{k}Y_{i}$, where 
$Y_{i} \sim \textup{Exp}(\mu)$, with $\mu=\left(\frac{1}{k}\sum_{i=1}^{k}\frac{1}{\mu_{i}}\right)^{-1}$. Define 
$d=S-Q$ and let use $\textup{Pr}\left[ I \right]$ to denote the occurrence probability of some event $I$. Then 
\begin{equation}
\label{eq:dProb}
\textup{Pr}\left [\left|d \right|\geq \delta^{2}\right] \leq \frac{k\sigma_{k}^{2}}{\delta^{2}},
\end{equation}
where  $\sigma^{2}_{k}$ is the sample variance of $\frac{1}{\mu_{i}}$, $i \in \left 
\{1,...,k \right \}$, defined as 
\begin{equation}
\label{eq:var}
\sigma^{2}_{k}=\frac{1}{k}\sum_{i=1}^{k}\frac{1}{\mu_{i}^{2}}-\left (\frac{1}{k}
\sum_{i=1}^{k}\frac{1}{\mu_{i}} \right)^{2}.
\end{equation}
\end{proposition}
\begin{IEEEproof}
It is known that for any random variable $X \sim \textup{Exp}(\mu)$, $\textup{E}\left[X \right]=\frac{1}{\mu}$ and 
$\textup{Var}\left[X \right]=\frac{1}{\mu^{2}}$. Therefore, $\textup{E}[S]=\sum_{i=1}^{k}\frac{1}{\mu_{i}}$ and 
$\textup{Var} \left[S\right]=\sum_{i=1}^{k} \frac{1}{\mu_{i}^{2}}$. Similarly, $\textup{E} \left[Q \right]=\frac{k}{\mu}$ and $\textup{Var}\left[Q \right]=\frac{k}{\mu^{2}}$, which, by the definition of $\mu$, yields $\textup{E}\left[Q \right]=
\sum_{i=1}^{k}\frac{1}{\mu_{i}}$ and $\textup{Var} \left[Q \right]=
\frac{1}{k}\left(\sum_{i=1}^{k}\frac{1}{\mu_{i}^{2}}\right)^{2}$. Then the Chebyshev inequality \cite{Papoulis03} yields
\begin{equation}
\begin{aligned}
\textup{Pr}\left [\left|d \right|\geq \delta^{2}\right]\leq \frac{1}{\delta^{2}}& 
\left(\sum_{i=1}^{k}\frac{1}{\mu_{i}^{2}}-\frac{1}{k}\left(\sum_{i=1}^{k}
\frac{1}{\mu_{i}} \right)^{2} \right)\\ 
\leq~&\frac{k\sigma_{k}^{2}}{\delta^{2}}. 
\end{aligned}
\end{equation}
\end{IEEEproof}
From Proposition \ref{Pr:Approx}, it can be concluded that for $\sigma_{k}^{2} \to 0$, it holds $P_{S}\approx P_{Q}$. In words, this can be described as follows. Let $X_{1},...,X_{k}$ be $k$ i.ni.d. exponential random variables. If their mean values are located near each other (i.e., if mean values exhibit small variance), then the probability distribution of their sum can be approximated by that of $k$ i.i.d. exponential random variables, say $Y_{1},...,Y_{k}$, with the identical mean being the average of mean values of $X_{1},...,X_{k}$. The approximation is shown in Fig. \ref{Approx} for four i.ni.d. exponential distributions, namely, $X_{1}\sim \textup{Exp}(3)$, $X_{2}\sim \textup{Exp}(4)$, $X_{3}\sim \textup{Exp}(6)$, and $X_{4}\sim \textup{Exp}(8)$. It can be seen that the proposed approximation performs very well, despite its simple form. 

Thus, provided that the variance of the random amount of harvested energy is relatively steady for $k_{m}$ arrivals, 
(\ref{eq:Energy}) can be approximated as 
\begin{equation}
\label{eq:EnergyT}
Y_{m}=\sum_{i=1}^{k_{m}}X'_{m,i},
\end{equation}
where $X'_{m,i} \sim \textup{Exp}(\mu_{m})$ for $i \in \{1,...,k_{m}\}$, and 
\begin{equation}
\label{eq:EnergyTT}
\mu_{m}=\left(\frac{1}{k_{m}}\sum_{i=1}^{k_{m}}\frac{1}{\mu_{m,i}} \right)^{-1}.
\end{equation}
Thus the probability density function of $Y_{m}$ can be approximated as that of a Poisson sum of i.i.d. exponential random variables.  
\subsection{Energy Consumption}
\label{subsec:enCons}
We assume that users arrive at every SBS according to some Poisson process. More precisely, every SBS $m \in \mathcal{M}$ is selected by $L_{m}$ users according to a Poisson process with rate $\alpha_{m}$, i.e., $L_{m}\sim 
\textup{Poi}(\alpha_{m})$. Thus the number of users that select SBS $m$ during a time interval of length $t$, denoted by 
$L_{m,t}$, follows a Poisson distribution with rate $\alpha_{m}t$. Note that at the end of inactive time, the total number of users that select an SBS $m \in \mathcal{M}$, shown by $L_{m,T}$, does not follow a Poisson distribution, since the inactive interval, $T_{m}$, is itself a random variable. We omit the subscript $m$ for the simplicity of the notation. Then, 
\begin{equation}
\label{eq:NumUser}
\begin{aligned}
f_{L_{T}}(l) &= \int_{t=0}^{\infty} \textup{Pr}\left[T=t\right]\left (f_{Y}(y) |T=t \right )dt\\ 
&=\int_{t=0}^{\infty} \frac{\left(1/\lambda \right)^k}{(k-1)!}t^{(k-1)}e^{-\frac{t}{\lambda}} 
\cdot \frac{(\alpha t)^{l}e^{-\alpha t}}{l!}dt\\
&= \frac{\alpha^{l}\lambda^{k}(l+k-1)!}{l!(k-1)!}\left(\lambda+\alpha \right )^{-(l+k)}.
\end{aligned}
\end{equation}

Without loss of generality, we assume that the required energy equals the transmission power. Therefore, 
by (\ref{eq:Energy}), for a user $n \in \mathcal{N}_{m}$ that requires a minimum transmission rate $r_{n,\min}$, the required energy, $q_{nm}=p_{nm}$, is calculated as 
\begin{equation}
\label{eq:UserEne}
q_{nm}=\frac{N_{0}+I_{nm}}{\left| h_{nm} \right|^{2}}\left(\exp(r_{m,\min})-1 \right).
\end{equation}
As described in Section \ref{sec:SysMod}, let $H_{nm}$ be a random variable following Rayleigh distribution with parameter $\frac{1}{\sqrt{2\beta_{nm}}}$. Then a random variable $X_{nm}=\left| H_{nm} \right|^{2}$ follows exponential distribution with parameter $\beta_{nm}$, i.e., $X_{nm}\sim \textup{Exp} (\beta_{nm})$. In the following, we omit subscripts $m$ and $n$ unless they are necessary to avoid ambiguity. Then, by (\ref{eq:Energy}) and due to the basic probability rule: 
$f_{Q}(q)=f_{X}(x)\left| \frac{dx}{dq}\right|$ for $q=g(x)$, the distribution of $Q$, denoted by $f_{Q}(q)$, follows as
\begin{equation}
\label{eq:EnePDF}
f_{Q}(q)=\left(\frac{\theta}{q^{2}}\right)\exp\left(-\frac{\theta}{q} \right),
\end{equation}
with $\theta= \beta \left(N_{0}+I\right)\left(\exp(r_{\min})-1\right)$, and we assume $r_{\min}$ is selected so that 
$\theta>0$.

Now, assume that user $n$ selects SBS $m$ at time $0<t+\delta t<T_{m}$, $\delta \approx 0$. As users are 
served on a first-come first-served basis, the already-consumed energy at SBS $m \in \mathcal{M}$, i.e., 
the amount of energy that is already allocated, is given by
\begin{equation}
\label{eq:ConEne}
Z_{nm}=\sum_{i=1}^{L_{m,t}}q_{im}.
\end{equation}
The distribution of $Q_{im}$ is given by (\ref{eq:EnePDF}), so that the exact distribution of $Z_{nm}$ can be calculated by using the Laplace transform; nonetheless, its exact distribution has a complicated form since the energy consumption of users, $Q_{im}$, are i.ni.d., yielding the Laplace transform of $Q$ to include the modified Bessel function. Moreover, its first and second moments do not exist. As a result, in order to make $Z_{nm}$ computationally traceable, we confine our attention to the worst-case scenario, where every user $n$ assumes that all prior users in the queue are allocated the maximum allowed energy $q_{m,\max}$; in other words, every user calculates the distribution of an upper-bound of the allocated energy. Therefore we redefine $Z_{nm}$ as
\begin{equation}
\label{eq:CoEneT}
Z_{nm}=\sum_{i=1}^{L_{m,t}}q_{m,\max}=L_{m,t}q_{m,\max}.
\end{equation}
Thus $Z_{nm}$ is uniquely defined by $L_{m,t}$, which is the number of arrivals in a Poisson process with rate $\alpha t$. That is,
\begin{equation}
\label{eq:CoEnPdf}
f_{Z}(z)= \frac{(\alpha t)^{z/q_{\max}} e^{-\alpha t}}{(z/q_{\max})!}.
\end{equation}
\textbf{Normal Approximation-} If $\alpha t$ is large enough, $f_{Z}(z)$ can be approximated by normal distribution. More precisely, for $\alpha t>1000$, $f_{Z}(z) \approx \textup{Nor}(q_{\max}\alpha t,q_{\max}^{2} \alpha t)$; in words, a Poisson distribution with rate $\alpha t$ is approximated by a normal distribution with mean and variance both equal to $\alpha t$. The normal approximation can be used already from $\alpha t>10$; however a correction factor should be included so that a good approximation is guaranteed.      

At SBS $m \in \mathcal{M}$ and for every user $n \in \mathcal{N}$, the residual energy is then calculated as 
\begin{equation}
\label{eq:ReEn}
R_{nm}=Y_{m}-Z_{nm}.
\end{equation}
The exact distribution of $R$ is calculated as follows:
\begin{equation}
\label{eq:ReEnPdfO}
\begin{aligned}
F_{R}(r)&=\textup{Pr}\left[R\leq r \right] \\ 
&=\sum_{z=0}^{\infty}\textup{Pr}\left [Y-z \leq r \right]f_{Z}(z)\\
&=\sum_{z=0}^{\infty}F_{Y}(z+r)f_{Z}(z).\\
\end{aligned}
\end{equation}
Thus,
\begin{equation}
\label{eq:ReEnPdfT}
\begin{aligned}
f_{R}(r)&=\sum_{z=\max \{0,-r\}}^{\infty}f_{Y}(z+r)f_{Z}(z)\\ 
&=\frac{e^{-\alpha t} \mu^{k}}{(k-1)!}\sum_{z=\max \{0,-r\}}^{\infty} 
\frac{(\alpha t)^{z/q_{\max}} (z+r)^{k-1}e^{-\mu(z+r)}}{(z/q_{\max})!}.\\
\end{aligned}
\end{equation}
The final expression of $f_{R}(r)$ in (\ref{eq:ReEnPdfT}) cannot be further simplified. Since this form is difficult to work with, we approximate $f_{R}(r)$ as follows. \\
\textbf{Normal Approximation-} We use the normal approximations of $Y$ and $Z$, described in Sections \ref{sec:energyModel} and \ref{subsec:enCons}, respectively. Then, if $Y_{m} \sim \textup{Nor}(\frac{k_{m}}{\mu_{m}}, \frac{k_{m}}{\mu_{m}^{2}})$ and $Z \sim \textup{Nor}(q_{\max}\alpha t , q_{\max}^{2}\alpha t)$, one concludes that $R \sim \textup{Nor}(\frac{k_{m}}{\mu_{m}}-q_{\max}\alpha t, \frac{k_{m}}{\mu_{m}^{2}}+q_{\max}^{2}\alpha t)$.
\section{User Association Problem}
\label{sec:Problem}
\begin{figure*}[t]
\centering
\includegraphics[width=0.90\textwidth]{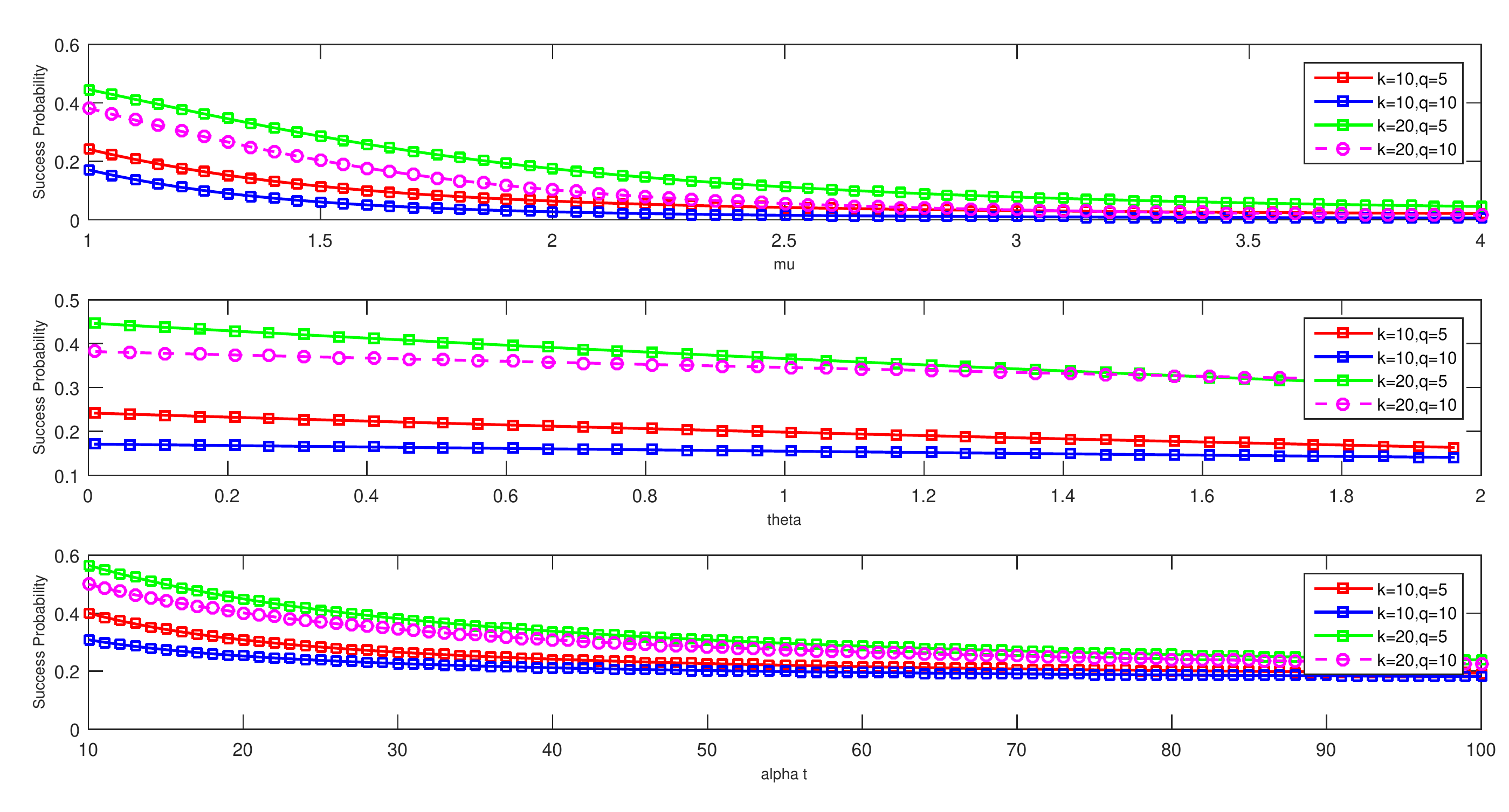}
\caption{Probability of successful selection as a function of inverse energy intensity ($\mu$), inverse channel quality ($\theta$), and SBS's user traffic ($\alpha$).}
\label{Fig:Success}
\end{figure*}

Upon arrival in the network, every user $n \in \mathcal{N}$ needs to select an SBS for the transmission of every packet. Thus, SBS selection is performed successively. In the rest of the paper, we call every round of selection as one 
\textit{trial}. As described in Section \ref{sec:SysMod}, we assume that users do not have any information about the channel qualities, as well as energy harvesting and user traffic profiles of small cells. Despite lack of knowledge, every user is interested in making successful decisions, as defined below. 
\begin{definition}[Successful Selection]
\label{de:SucSel}
A selection is successful if the following two conditions are satisfied
\begin{itemize}
\item $q_{nm}\leq q_{m,\max}$, and
\item $R_{nm}\geq q_{nm}$.
\end{itemize}
\end{definition}
In words, at every transmission, desired is to select an SBS for which: i) The required energy to guarantee the desired QoS is less than the maximum allowed energy; and ii) By the time of selection, the residual energy at the SBS is larger than the required energy. Thus, for every user $n$, the success probability when connecting to SBS $m$, denoted by 
$p_{nm,s}$, is given by
\begin{equation}
\label{eq:SucPro}
p_{nm,s}=\int_{q=0}^{q_{\max}}\int_{r=q}^{\infty}f_{R}(r)f_{Q}(q) dq dr,
\end{equation}
and the failure probability yields $p'_{nm,s}=1-p_{nm,s}$. The integral in (\ref{eq:SucPro}) cannot be calculated in 
closed-form, even if normal approximation is used for $f_{R}(r)$. Nonetheless, given $q_{\max}$, $p_{nm,s}$ can be calculated numerically. In order to derive an explicit formula, one approach would be to develop a lower-bound for the success probability by requiring that $R_{nm} \geq q_{m,\max}$, so that the dependency on $q$ is eliminated. Formally,   
\begin{equation}
\label{eq:SucProUp}
\begin{aligned}
p_{nm,s} & \geq \textup{Pr}\left [R_{nm}\geq q_{m,\max} \right]\cdot \textup{Pr}\left[ q_{nm}<q_{m,\max} \right] \\ 
& = (1-F_{R}(q_{m,\max})) F_{Q}(q_{m,\max}),
\end{aligned}
\end{equation}
where for any random variable $X$, $F_{X}(x)$ denotes the cumulative density function. Then, by using the normal approximation of the first term in the right-hand-side of (\ref{eq:SucProUp}), $p_{nm,s}$ is approximated as
\begin{equation}
\label{eq:SucProApF}
\begin{aligned}
p_{nm,s}(\mu,\alpha,\theta,t)\geq \hspace{100pt}&\\
\left(\frac{1}{2}-\frac{1}{2}\textup{erf}\left(\frac{q_{m,\max}-\frac{k_{m}}{\mu_{m}}+q_{m,\max}\alpha_{m} t}{\sqrt{2}
\left(\frac{k_{m}}{\mu_{m}^{2}}+q_{m,\max}^{2}\alpha_{m}t \right)}\right) \right) & \exp \left(\frac{-\theta_{nm}}{q_{m,\max}} \right),\\
\end{aligned}
\end{equation}
where $\textup{erf}(\cdot)$ is the error function. Fig. \ref{Fig:Success} depicts the success probability as a function of involved parameters. Note that according to our system model, the intensity of energy arrivals is inversely proportional to $\mu$, and $\theta$ is inversely related to channel quality. As expected, the figure shows that the success probability decreases with increasing $\mu$ and $\theta$. Similarly, it decreases with increasing $\alpha$ (or $\alpha t$), which is directly related to the number of users in the queue, i.e., the already-allocated (consumed) energy at the time the SBS is selected by the user. Moreover, for some fixed $k$ and $\alpha$ (which determine the duration of inactive step and queue length), smaller $q_{\max}$ increases the success probability, since for every user, smaller $q_{\max}$ results in smaller amount of already-consumed energy, so that a larger number of users can be served. Note that selecting $q_{\max}$ very small would also have an adverse effect, since many users with weak channels cannot meet the required QoS, although larger number of users are served. Similarly, for fixed $q_{\max}$ and $\alpha$, larger $k$ results in higher success probability, since it implies that more energy is stored during the inactive mode. Note that choosing $k$ too large results in delayed services. 
%
 
For every user $n \in \mathcal{N}$, the number of packets to be transmitted is denoted by $J_{n}$, which we assume is large enough. Moreover, based on our previous discussion, for every user $n \in \mathcal{N}$, when selecting each SBS 
$m \in \mathcal{M}$, the reward can be regarded as a Bernoulli random variable with success probability 
$p_{nm,s}(\cdot)$, which is lower-bounded as given in (\ref{eq:SucProApF}). Thus, at every selection (transmission) round $j=1,...,J_{n}$, we define the following utility (reward) function for user $n$, if some SBS $m \in \mathcal{M}$ is selected:
\begin{equation}
\label{eq:Utility}
u_{n,j}(m)=\begin{cases}
1 & \textup{if} \hspace{5pt} r_{nm}\geq r_{n,\min} \\ 
0 & \textup{otherwise (o.w.)} 
\end{cases}.
\end{equation}
Since SBSs are not synchronized, and the duration of active and inactive steps are random, at each trial 
$j=1,...,J_{n}$, only a set of SBSs $\mathcal{M}_{j} \subseteq \mathcal{M}$ is available. Now, let $\mathcal{O}$ be the set of all SBS selection strategies (decision making policies). Moreover, assume that user $n \in \mathcal{N}$ selects some SBS $m_{n,j}^{(\sigma)}$ at each step $j$, according to some selection policy $\sigma \in \mathcal{O}$, which results in some (instantaneous) utility $u_{n,j}\left(m_{n,j}^{(\sigma)} \right)$. Then the (accumulated) utility of policy $\sigma$ yields
\begin{equation}
\label{eq:UtilitySigma}
U_{n,\sigma}=\sum_{j=1}^{J_{n}}u_{n,j}\left(m_{n,j}^{(\sigma)}\right).
\end{equation}
In words, the reward of the policy is the accumulated reward achieved by selecting actions suggested by that policy over the entire transmission horizon. Ideally, every user $n$ wants to use some policy $\sigma \in \mathcal{O}$ so as to solve the following optimization problem in order to obtain the achievable reward of the best selection policy: 
\begin{equation}
\label{eq:OptProb}
\underset{\sigma \in \mathcal{O}}{\textup{maximize}} \hspace{5pt} U_{n,\sigma}.
\end{equation}
 In order to solve problem (\ref{eq:OptProb}), every user faces the following difficulties: i) Statistical information of energy harvesting, channel quality and SBS traffic is not available; ii) Success probability is time-varying: it depends on the length of time interval beginning at the time an SBS enters the inactive mode until it is selected by the user; iii) The set of available SBSs (i.e., those that can be selected for transmission) varies at every transmission round, since the length of inactive and active steps are non-deterministic. As a result, the solution is infeasible and the user might revert to a less ambitious goal. 

Let $\mathfrak{O}$ be an ordering (permutation) of $M$ SBSs. We use $\mathfrak{O} \left(\mathcal{M}_{j} \right)$ to denote the best choice in $\mathcal{M}_{j}$ that is highest ranked in $\mathfrak{O}$. Also, an \textit{$\mathfrak{O}$-policy} corresponding to the ordering $\mathfrak{O}$ is the policy that selects, at each time trial $j$, the action $\mathfrak{O} \left(\mathcal{M}_{j} \right)$, i.e., the available action that is highest ranked by $\mathfrak{O}$ \cite{Kleinberg10:RBS}. If user $n \in \mathcal{N}$ uses policy $\mathfrak{O}$, the reward is given by
\begin{equation}
\label{eq:Utility}
U_{n,\mathfrak{O}}= \sum_{j=1}^{J_{n}} u_{n,j}(\mathfrak{O} \left(\mathcal{M}_{j} \right)).
\end{equation}
Moreover, for every $n$, by $\mathfrak{O}^{*}$ we denote an ordering that solves the optimization problem in (\ref{eq:OptProb}), i.e., the best ordering, that yields a reward $U_{n,\mathfrak{O}^{*}}:= U_{n}^{*}$. Then the \textit{regret} of any selection policy $\sigma$ is defined as 
\begin{equation}
\label{eq:regret}
d_{n,\sigma}= \mathbb{E}\left[U_{n}^{*}-\sum_{j=1}^{J_{n}}u_{n,j}\left(m_{n,j}^{\sigma}\right) \right],
\end{equation}
where $\mathbb{E}\left[\cdot \right]$ is the mathematical expectation which is taken with respect to the random choices of the algorithm as well as the randomness in the utility function. In words, the regret of an algorithm is defined as the expected difference between the accumulated utility achieved by that algorithm and the maximum achievable utility. Then, every user $n$ opts to minimize the regret, i.e., to solve the following optimization problem:
\begin{equation}
\label{eq:regretMin}
\underset{\sigma \in \mathcal{O}}{\textup{minimze}}\hspace{5pt}d_{n,\sigma}.
\end{equation}
In the next section we show that problem (\ref{eq:regretMin}) can be cast and solved as an adversarial multi-armed bandit game with sleeping arms. 
\begin{remark}
\label{re:action}
In case a user $n \in \mathcal{N}$ intends to select a set $\mathcal{M}_{n} \subseteq \mathcal{M}$ of SBSs with cardinality $M_{n}>1$, every combination of $M_{n}$ out of $M$ SBSs is regarded as a \textit{multi-SBS} or 
\textit{super-SBS}. That is, a set of multi-SBSs is defined as $\mathcal{M}'=C\left(M, M_{n}\right)$, with its cardinality being $M'=\binom{M}{M_{n}}=\frac{M!}{(M-M_{n})!M_{n}!}$. Consider a multi-SBS $m' \in \mathcal{M}'$ that consists of 
$M_{n}$ SBSs labeled as $1,...,M_{n}$. Let $\mathbb{I}_{i}$ denote the availability of any SBS or multi-SBS $i$, so that 
$\mathbb{I}_{i}=1$ if $i$ is available (inactive mode) and $\mathbb{I}_{i}=0$ otherwise. Then the availability of 
multi-SBS $m'$ is defined as
\begin{equation}
\label{eq:AvailMult}
\mathbb{I}_{m'}=\prod_{i=1}^{M_{n}}\mathbb{I}_{i},
\end{equation}
which means that a multi-SBS $m'$ is available only if all of its included SBSs are available. Moreover, the achieved utility through multi-SBS $m'$ yields
\begin{equation}
\label{eq:RewMult}
u_{n,j}(m')=\sum_{i=1}^{M_{n}}u_{n,j}(i),
\end{equation}
that is, the reward of every multi-SBS is the aggregate reward of its individual components.
\end{remark}
\section{Bandit-Theoretical Model and Solution}
\label{sec:Bandit}
Multi-armed bandit is a class of online optimization problems, where an agent, given no prior information, selects an arm from a finite set of arms in successive trials. Upon being pulled, every arm produces some reward, which is drawn from the reward generating process of that arm. The agent observes only the reward of the played arm and not those of other arms. 
Bandits can be classified based on the reward generating process of arms. For instance, in \textit{adversarial bandits}, the instantaneous rewards of arms cannot be attributed to a specific probability distribution; that is, the reward generating processes vary adversarially. In \textit{stochastic bandits}, however, rewards can be attributed to a specific probability distribution. As a result of lack of prior information, at each trial, the agent may choose some inferior arm in terms of reward, yielding some regret that is quantified by the difference between the reward that would have been achieved had the agent selected the best arm and the actual achieved reward. The agent intends to decide which arm to pull in a sequence of trials so that its accumulated regret over the game horizon is minimized. This problem is an instance of exploration-exploitation dilemma, i.e., the tradeoff between taking actions that yield immediate large rewards on the one hand and taking actions that might result in larger reward only in future, for instance activating an inferior arm only to acquire information, on the other hand. A solution of a bandit problem is thus a decision making strategy called policy or allocation rule, which determines which arm should be played at successive rounds so that the optimal balanced between exploitation and exploration is achieved. While in most bandit problems all arms are available during the entire horizon, in \textit{sleeping bandits}, the set of available arms is time-varying, so that at each trial the arm to be pulled is selected from a subset of arms. Similar to the reward process, the availability can be adversarial or stochastic. In case of limited availability, at every trial, the agent tries to pull the best arm with respect to the ordering of available arms.   
\subsection{Bandit-Theoretical Model of User Association}
\label{subsec:BanditModel}
According to our system model and problem formulation, the user association problem can be modeled and solved by using 
\textit{adversarial sleeping bandit} model. In this model, every user $n \in \mathcal{N}$ is an agent, whereas every SBS 
$m \in \mathcal{M}$ represents an arm, whose reward generating process is a Bernoulli random variable with time-varying parameter. Using such model can be justified by the following reasons:
\begin{itemize}
\item Each SBS is available to be selected by users as soon as the active (transmission) step comes to an end and the inactive (energy harvesting) step begins. Without loss of generality, we assume that transmissions are performed sequentially and one unit of time is spent for every user. Thus, for every SBS $m \in \mathcal{M}$, the duration of transmission step depends on the total number of users that have selected that specific SBS, as given 
by (\ref{eq:NumUser}). As a result, the SBS availability is stochastic. 
\item The utility of every user upon selecting any SBS is a Bernoulli random process with a time-varying success probability. Consequently, the utility can be considered adversarial. 
\item Users do not have any prior information on the success probability of selecting each one of SBSs.
\item After selecting an SBS, the user only observes whether the transmission via that specific SBS has been successful or not. No other information is revealed.
\end{itemize}
%
\subsection{Algorithmic Solution}
\label{subsec:Solution}
We use algorithm EXP4 \cite{Auer03:NMB} for sleeping bandits as suggested in \cite{Kleinberg10:RBS}. At each round 
$j \in J_{n}$, the algorithm assigns some selection probability $\textup{Pr}[m]=a_{nm,j}$ to each arm $m \in \mathcal{M}$, so that $\sum_{m=1}^{M} a_{nm,j}=1$. To calculate $\mathbf{a}_{n,j}=\left(a_{n1,j},..., a_{nM,j}\right)$, the algorithms relies on ($M!$+1) experts: one of them being the uniform expert that corresponds to the uniform distribution over $M$ arms 
($a_{nm,j}=\frac{1}{M}$, for all $m \in \mathcal{M}$), and each one of the other $M!$ experts corresponds to one ordering 
$\mathfrak{O}$ that advices to select the arm with the highest rank that is available. The algorithm weighs the past performance of each expert exponentially, calculates $\mathbf{a}_{n,j}$ by combining weighted experts and selects an action using $\mathbf{a}_{n,j}$. The procedure is summarized in \textbf{Algorithm \ref{Alg:EXP4}}, where we omit subscript $n$ for simplicity. Details can be found in \cite{Auer03:NMB} and \cite{Kleinberg10:RBS}. It should be mentioned that in our problem setting, the adversary that selects the rewards (or losses) is non-oblivious (adaptive), due to the following reason: The actions of each user yields higher traffic to its selected SBSs, which in turn impacts the selections of other users as they learn some SBSs have higher traffic load. Their decisions then impact the initial user, and so on. Thus we need to know the regret against an adaptive (non-oblivious) adversary, as stated in Theorem \ref{th:RegEXPTwo}. 
It is worth noting that \textbf{Algorithm \ref{Alg:EXP4}} keeps track of $M!+1$ weights as its bottleneck, resulting in space and time complexity of $O(M!+1)$.    
\begin{remark}
\label{re:AlgMulti}
\textbf{Algorithm \ref{Alg:EXP4}} also works for the case where some users are willing to select multiple SBSs. In fact, it is enough to use $\mathcal{M}'$ instead of $\mathcal{M}$ as the action set, as discussed in Remark \ref{re:action}. Note that in this case the space and time complexity are of $\Omega \left( (M^{M_{n}})! \right)$ knowing that $C(M,M_{n})=\Omega \left(M^{M_{n}} \right)$.
\end{remark}
\begin{algorithm}
\caption{EXP4-SB \cite{Kleinberg10:RBS}}
\label{Alg:EXP4}
\small
\begin{algorithmic}[1]
\STATE Select $\gamma \in (0,1]$;
\STATE Label $M!+1$ experts by integer values $1,...,M!+1$;
\STATE Initialize $w_{k,1}=1$ for $k=1,...,M!+1$;
\FOR{$j=1,...,J$}
\STATE For uniform expert, let $\mathbf{b}_{j}^{(1)}=\left(\frac{1}{M},...,\frac{1}{M}\right)$.
\STATE For each one of the other $M!$ experts, i.e., for the $k$-th ordering $\mathfrak{O}$, $b_{m,j}^{(k)}$, 
$k=1,...,M!$, is defined as 
\begin{equation}
\label{eq:ProbExpert}
b_{m,j}^{(k)}=
\begin{cases}
 1, & m=\mathfrak{O}(\mathcal{M}_{j})\\ 
 0, & \text{o.w}  
\end{cases},
\end{equation}
and $\mathbf{b}_{j}^{(k)}=\left(b_{1,j}^{(k)},...,b_{M,j}^{(k)}\right)$.
\STATE Let $W_{j}=\sum_{k=1}^{M!+1}w_{k,j}$.
\STATE For every arm $m \in \mathcal{M}$, calculate the selection probability as 
\begin{equation}
\label{eq:selProb}
a_{m,j}=(1-\gamma)\sum_{k=1}^{M!+1}\frac{w_{k,j}b_{m,j}^{(k)}}{W_{j}}+\frac{\gamma}{M}.
\end{equation}
\STATE Select an action $m_{j}$ according to probability distribution $\mathbf{a}_{m}=
\left(a_{1,j},...,a_{M,j}\right)$.
\STATE Play and observe the reward $u_{j}(m_{j})$.
\STATE For every action $m' \in \mathcal{M}$ set
\begin{equation}
\label{eq:RewEst}
\hat{u}_{j}(m')=
\begin{cases}
\frac{u_{j}(m_{j})}{a_{m',j}}, & m'=m_{j}\\ 
0, & \text{o.w}  
\end{cases}.
\end{equation}
\STATE For every expert $k=1,...,M!+1$ set
\begin{equation}
\label{eq:ExpUpdateO}
\hat{y}_{k,j}=\mathbf{b}_{j}\cdot \hat{\mathbf{u}}_{j}
\end{equation}
\begin{equation}
\label{eq:ExpUpdateT}
w_{k,j+1}=w_{k,j}\exp \left(\frac{\gamma \hat{y}_{k,j}}{M} \right)
\end{equation}
\ENDFOR
\end{algorithmic}
\end{algorithm}
%
\begin{theorem}
\label{th:RegEXPTwo}
With high probability, Algorithm \ref{Alg:EXP4} achieves a regret of $O\left(M\sqrt{J_{n}\log(M)}+\sqrt{J_{n}}\right)$ with respect to the best ordering, against an non-oblivious adversary.
\end{theorem}
\begin{IEEEproof}
By Theorem 15 of \cite{Kleinberg10:RBS}, Algorithm \ref{Alg:EXP4} achieves a regret of 
$O \left(M\sqrt{J_{n}\log(M)} \right)$ with respect to the best ordering against an oblivious adversary. On the other hand, by Lemma 4.1. of \cite{Bianchi06:PLG}, if the expected regret of any policy against an oblivious adversary is bounded by some constant $B$, then for all $\delta>0$ and with probability at least $1-\delta$, its actual accumulated regret against a non-oblivious adversary is bounded by $B+\sqrt{\frac{J_{n}}{2} \log\left(\frac{1}{\delta} \right)}$. Therefore, the proof follows.

\end{IEEEproof}
\section{Numerical Analysis and Discussions}
\label{sec:Numeric}
We consider a small cell network with $M=5$ small cells. Parameters of small cells, as discussed through the paper, are gathered in Table \ref{Tb:SCPar}.\footnote{Parameters are selected at random.} 
\begin{table}[ht]
\caption{Simulation Parameters}
\label{Tb:SCPar}
\begin{center}
  \small
  \begin{tabular}{|c|c|c|c|c|c|}
  \hline
   \backslashbox{Small Cell}{Parameter}  & $\lambda_{m}$ & $\mu_{m}$ &$\alpha_{m}$ & $q_{m,Max}$& $k_{m}$ \\ \hline 
   1 & 80  & 0.03 & 10 & 7 & 50   \\ \hline
	 2 & 70  & 0.06 & 12 & 8 & 100  \\ \hline
	 3 & 80  & 0.09 & 10 & 9 & 69   \\ \hline
	 4 & 130 & 0.12 & 15 & 6 & 40   \\ \hline
	 5 & 120 & 0.11 & 10 & 7 & 40   \\ \hline 
  \end{tabular}
 \end{center} 
\end{table}  

In the following, we show the decision making behavior of two \textit{exemplary} users. Note that the actual number of users in the network varies randomly as described in Section \ref{subsec:enCons}: we use these two users only as examples to clarify the decision making process and to investigate its performance. Let the Hadamard (element-wise) product of two matrices $\mathbf{A}$ and $\mathbf{B}$ be denoted as $\mathbf{A}{\circ}\mathbf{B}$. Furthermore, by 
$\mathbf{A^{\circ2}}$ we denote the element-wise squared of matrix $\mathbf{A}$. In addition, for every matrix 
$\mathbf{A}$, $\mathbf{A}\left[n,m\right]$ stands for the element located at $n$-th row and $m$-th column.
As conventional, we assume that the channel gain matrix $\mathbf{H^{\circ2}}$ can be written as $\mathbf{H^{\circ2}}=\mathbf{F}{\circ}\mathbf{G}$, where $\mathbf{F}$ and $\mathbf{G}$ are average fading gain and path-loss matrices, respectively. Then $\mathbf{G}\left[n,m\right]$ and $\mathbf{F}\left[n,m\right]$ correspondingly denote the average fading gain and path-loss of the link between user $n$ and small cell $m$, for $n \in \{1,2\}$ and $m \in \{1,2,3,4,5\}$. We let 
$\mathbf{F}=\begin{bmatrix}
0.90 & 0.20 & 0.10 & 0.10 & 0.20\\ 
0.05 & 0.05 & 1.00 & 0.30 & 0.30
\end{bmatrix}$ 
and
$\mathbf{G}=\begin{bmatrix}
0.80 & 0.10 & 0.05 & 0.20 & 0.10\\ 
0.10 & 0.20 & 1.00 & 0.10 & 0.02
\end{bmatrix}$. Moreover, we select 
$J_{n}=5 \times 10^{4}$, $r_{n,\min}=0.5$, and $\mathbf{I}_{nm}=\begin{bmatrix}
1.0 & 3.0 & 2.0 & 4.0 & 2.0
\end{bmatrix}$ for $n \in \{1,2\}$. Moreover, $N_{0}=1$ and the exploration parameter equals $\gamma=0.05$. 

\begin{figure}[ht]
\centering
\subfigure[User 1]{
\includegraphics[width=0.45\textwidth]{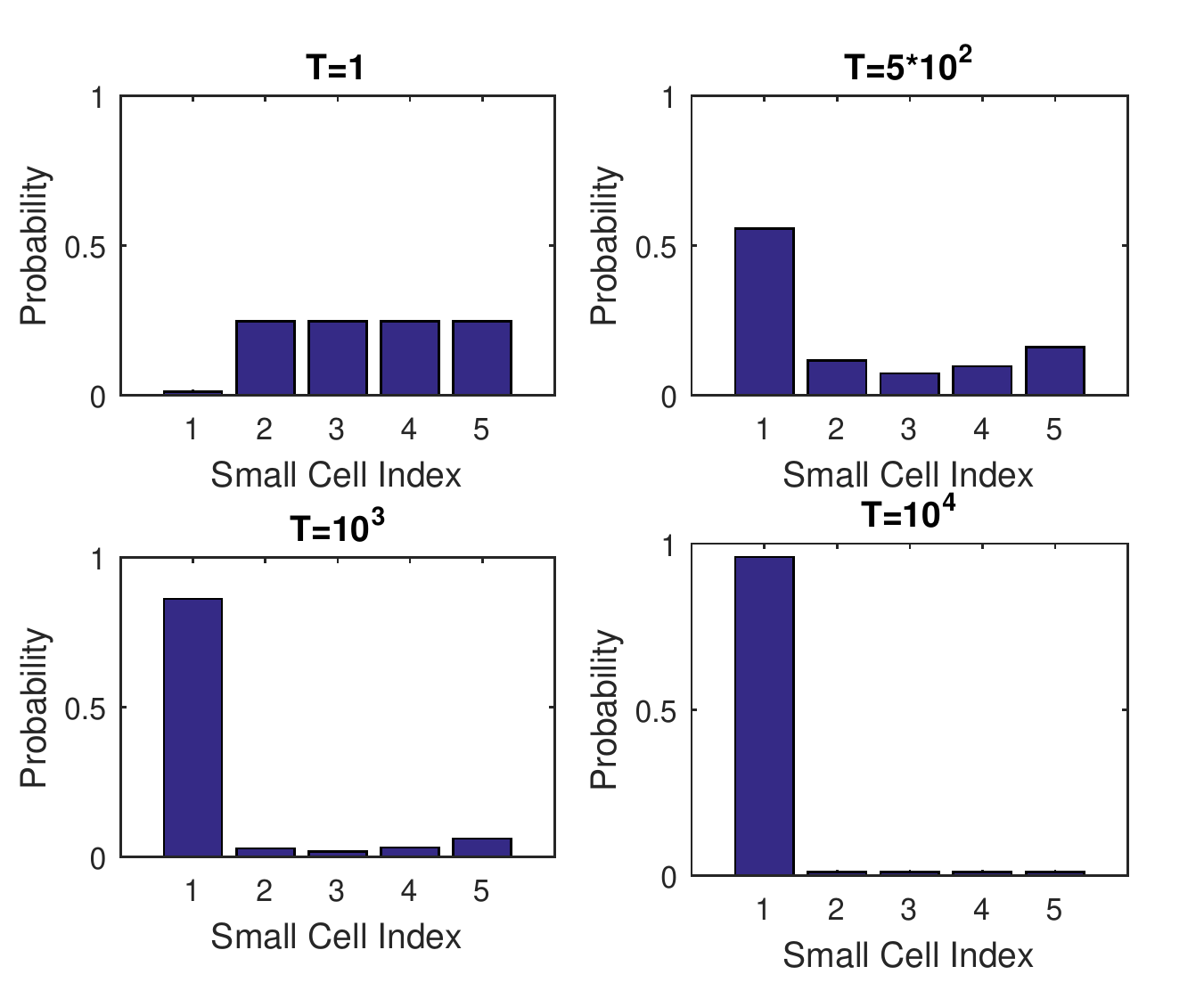}
\label{Fig:MSOne}}
\quad
\subfigure[User 2]{
\includegraphics[width=0.45\textwidth]{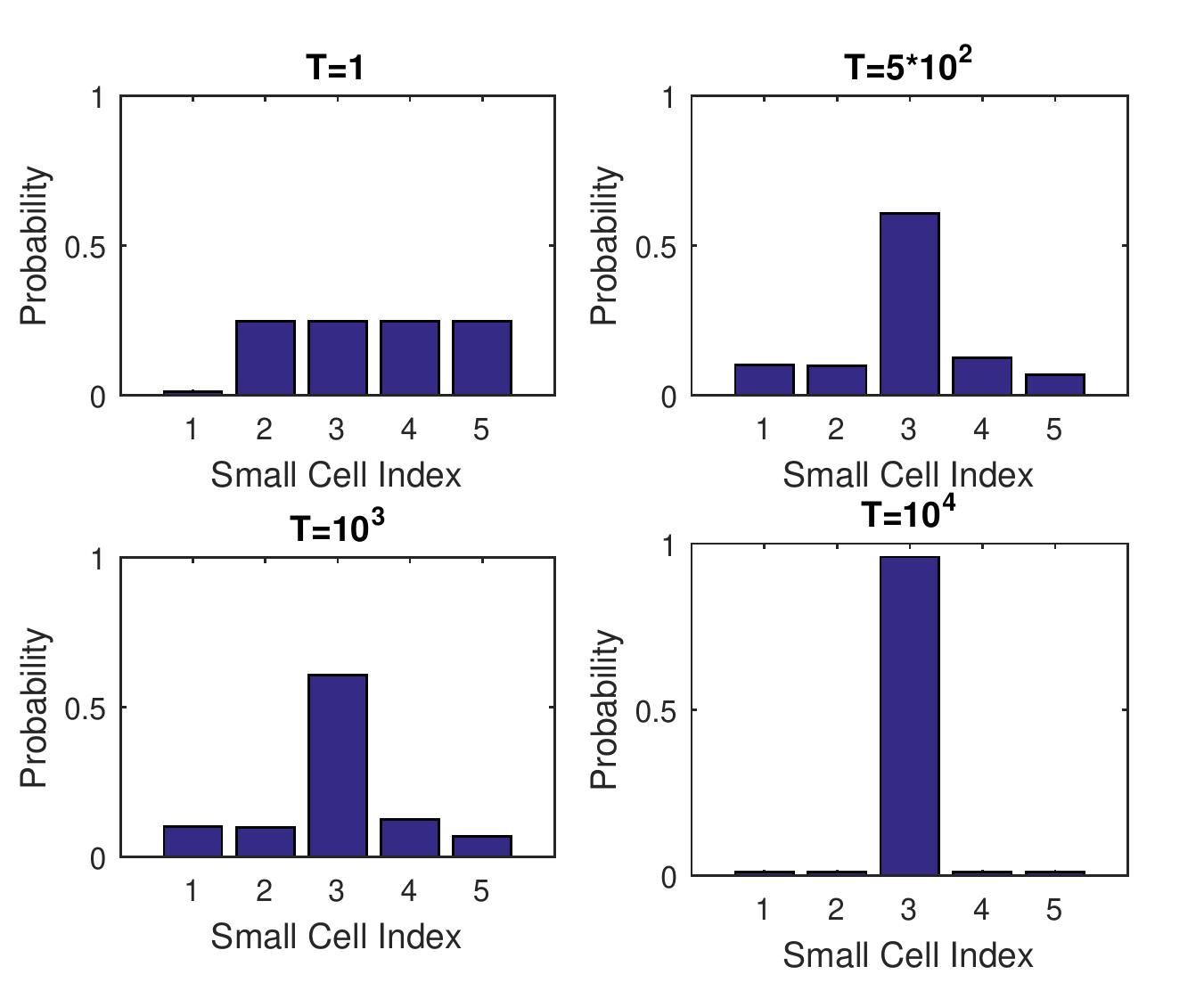}
\label{Fig:MSTwo}}
\caption{Evolution of mixed strategies for two exemplary users.}
\label{Fig:MS}
\end{figure}

\begin{figure}[t]
\centering
\includegraphics[width=0.45\textwidth]{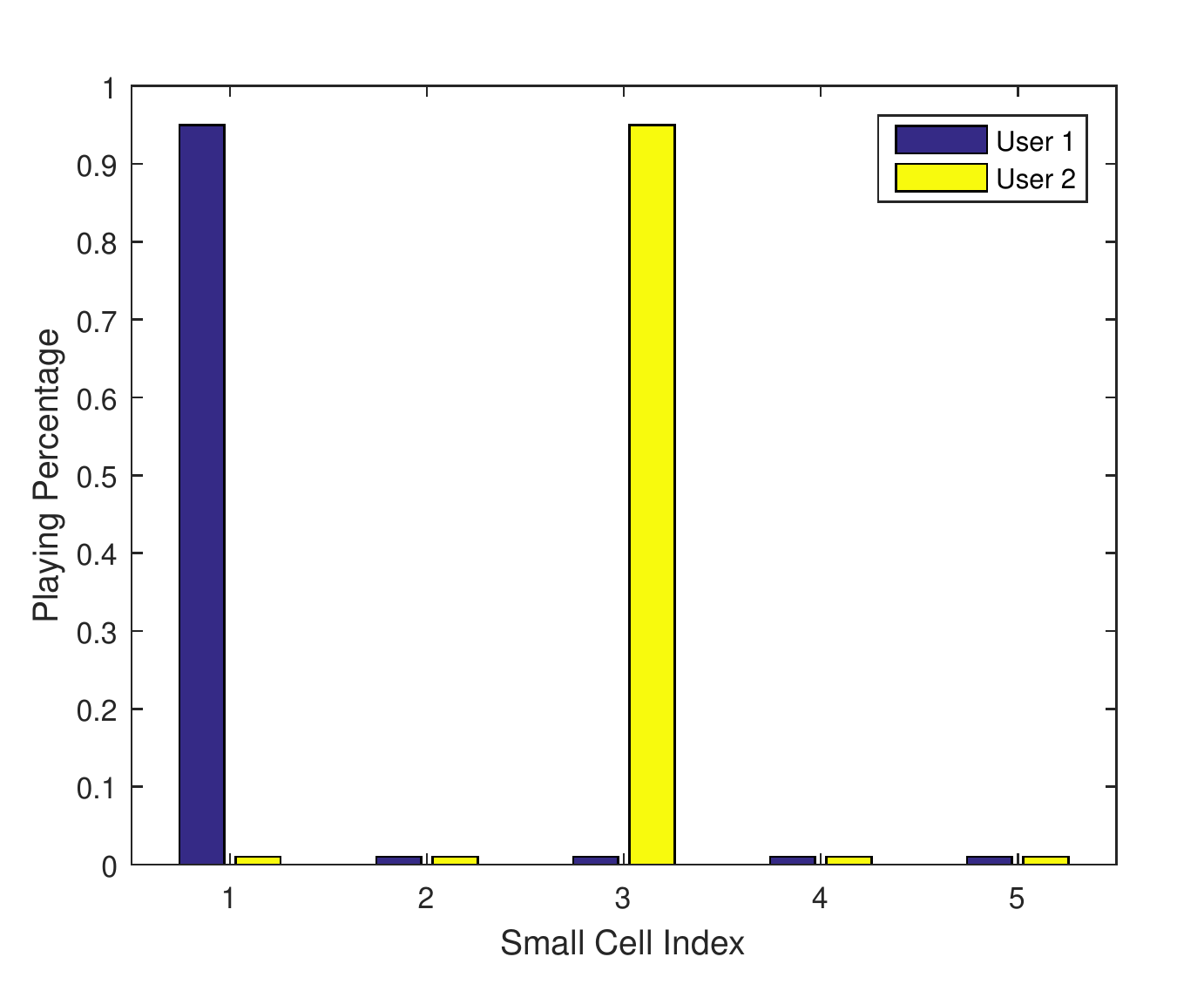}
\caption{The fraction of time (in percent) each small cell is selected by users.}
\label{Fig:Perc}
\end{figure}

Fig. \ref{Fig:MS} illustrates the evolution of the mixed strategies of the two typical users, namely, User 1 and 
User 2. The percentage of time each small cell is selected by these two users is shown in Fig. \ref{Fig:Perc}. 

\begin{figure}[t]
\centering
\includegraphics[width=0.45\textwidth]{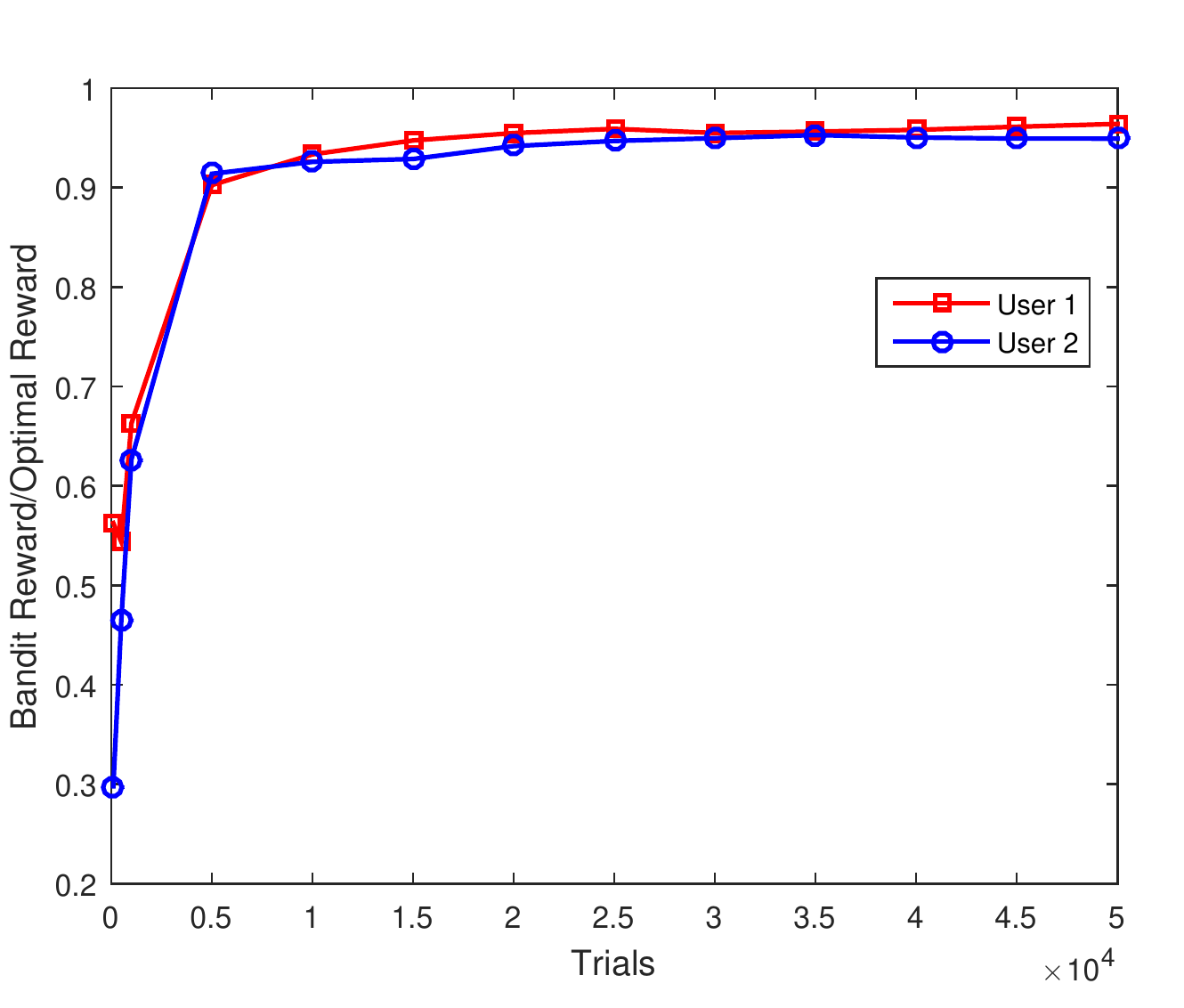}
\caption{The achieved reward of the bandit model compared to optimal selection.}
\label{Fig:Rew}
\end{figure}

Fig. \ref{Fig:Rew} shows the reward achieved by the bandit approach compared to the best fixed choice that has the highest success probability (optimal SBS, selected through exhaustive search given all information), as discussed in Section 
\ref{sec:Problem}. 

From Figs. \ref{Fig:MS}, \ref{Fig:Perc}, and \ref{Fig:Rew}, it can be concluded that the mixed strategy converges to the optimal choice in the sense of maximum success probability and that the best small cell is played almost all the time, so that the average performance converges to that of optimal selection given full statistical information of channel quality and network characteristics.  

\begin{figure}[t]
\centering
\includegraphics[width=0.45\textwidth]{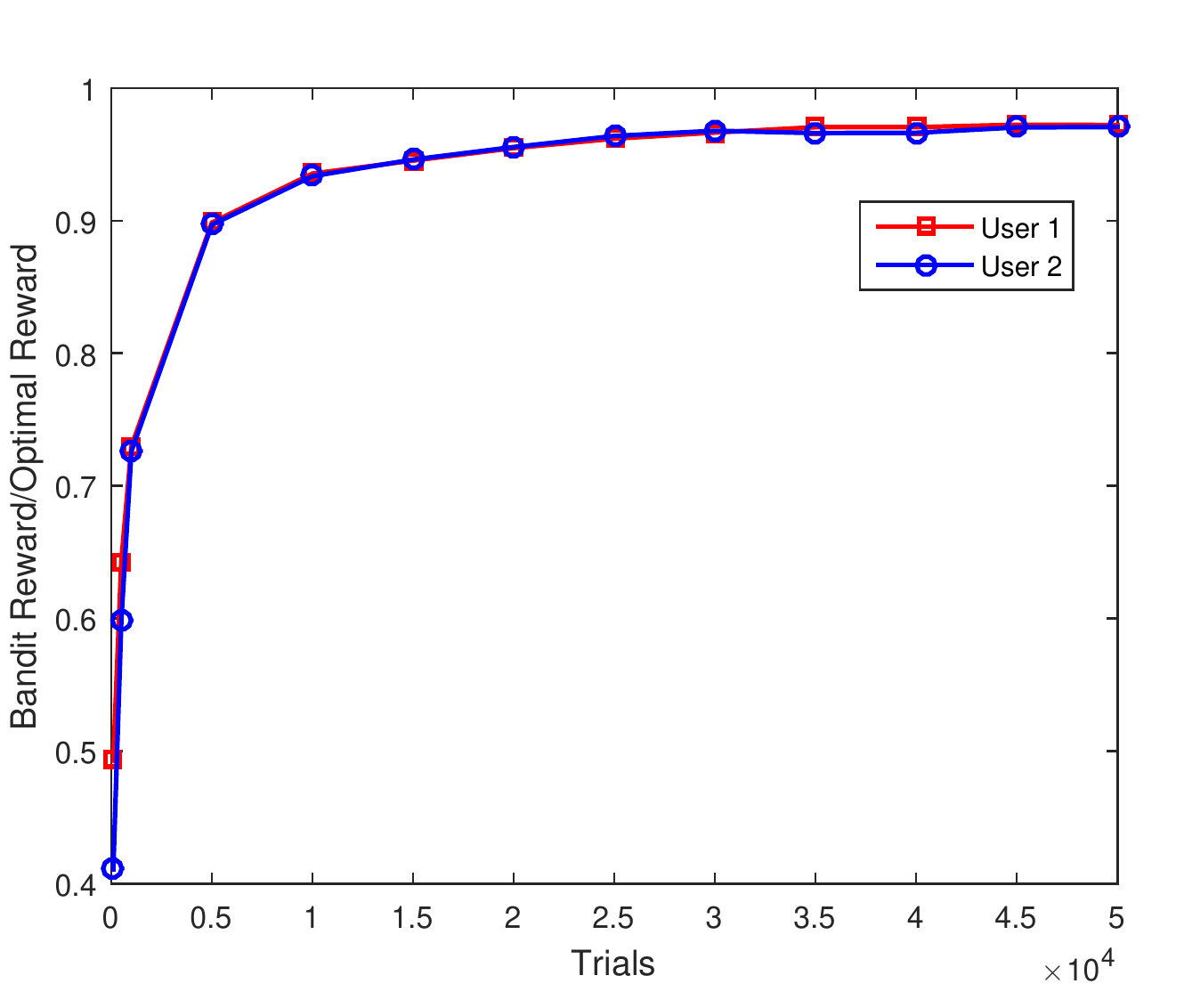}
\caption{The achieved reward of the bandit model compared to optimal selection, multiple SBSs are selected at every transmission round.}
\label{Fig:RewMul}
\end{figure}

In the next step, we assume that every user intends to select two SBSs at every trial out of the first four SBSs in Table \ref{Tb:SCPar}. As a result, the new action set, $\mathcal{M}'$, consists of $M'=6$ super-actions each including two actions, namely, $\mathcal{M}'= \left\{(1,2),(1,3),(1,4),(2,3),(2,4),(3,4) \right\}$. The performance compared to the optimal is shown in Fig. \ref{Fig:RewMul}. Mixed strategies and selected actions are similar to 
Figs. \ref{Fig:MS} and \ref{Fig:Perc}, hence are omitted.

\begin{figure}[t]
\centering
\includegraphics[width=0.49\textwidth]{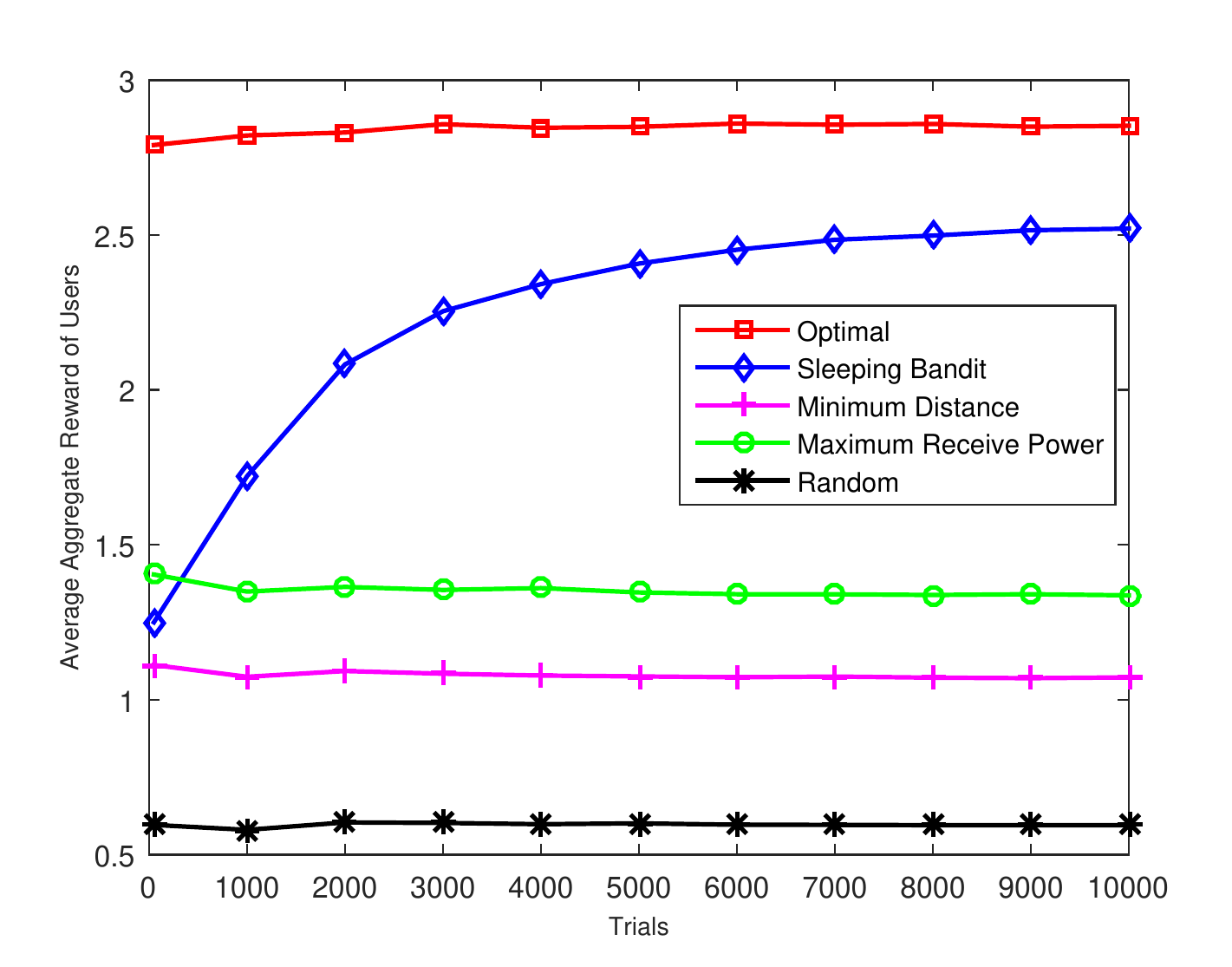}
\caption{Performance Comparison.}
\label{Fig:CompL}
\end{figure}

Finally, in Fig. \ref{Fig:CompL}, we consider a large network consisting of $M=8$ small cells, and we investigate the aggregate performance of a set $\mathcal{W} \subset \mathcal{N}$ of users with cardinality $W=10$. Once again, note that for each trial, the total number of users $N$ varies randomly as described in Section \ref{subsec:enCons}; From this point of view, $\mathcal{W}$ represents a set of users under investigation. We assume that each user transmits for 
$J_{w}=10^{4}$ trials, but they are not synchronized. For comparison, we also evaluate few other user association schemes that are widely-used, as described below.   
\begin{itemize}
\item Optimal Assignment: In this scenario, every user (or a central unit) is provided with \textit{complete} statistical information of energy harvesting, user arrival and channel qualities at every small cell. Moreover, other characteristics of small cells, including $q_{m,\max}$ and $k_{m}$, $m \in \mathcal{M}$, are known. Given information and through exhaustive search, every user is assigned to the small cell that offers the highest success probability.
\item Maximum Received Power Assignment: In this scenario, user association is performed by a central unit given average channel gain matrix, $\mathbf{H}$. The (statistical) information of energy harvesting and user arrival remain unknown. By means of exhaustive search, every user is assigned to the SBS to which it has the maximum average channel gain. Assignment based on received power has been widely used to solve the user association problem (e.g., in \cite{Lin15:OUA}).
\item Nearest SBS (Minimum Distance) Assignment: In this scenario, user association is performed by a central unit given geographical locations of users and SBSs, as well as the path-loss exponent. In our model, we assume that the path-loss exponent is equal for all links; thus larger distance yields larger path-loss and \textit{vice versa}. By means of exhaustive search, every user is assigned to the SBS to which it has the minimum path-loss. It is clear that the performance of maximum received power method serves as an upper-bound for that minimum distance assignment. Distance-based assignment is a conventional method to solve different types of association problems (e.g., in \cite{Son11:BSO}).
\item Sleeping Bandit Assignment: In this scenario, the proposed bandit model and algorithm is used for distributed user association given no information.
\item Random Assignment: Users are associated randomly.
\end{itemize}
In Fig. \ref{Fig:CompL}, it can be seen that the bandit algorithm exhibits superior performance compared to conventional assignment approaches such as maximum received power and minimum distance assignment, although those methods require channel and/or path-loss information. In fact, conventional methods are mostly unable to combat the uncertainty hidden in energy harvesting. As the final remark, it should be mentioned that not all user association methods can be directly compared to each other. This is because, as discussed in Section \ref{sec:Introduction}, every method is designed for a specific system model and aims at optimizing a particular performance metric. 
\section{Conclusion}
\label{sec:Conclusion}
We have proposed a new probabilistic framework to model energy harvesting in wireless small cell networks. We then investigated the distributed user association problem in such networks while taking the uncertainty into account, which is caused by various random effects of multi-user energy harvesting networks, including non-deterministic energy arrival as well as non-deterministic energy consumption. We proposed a bandit framework to efficiently solve the user association problem in a distributed manner where users suffer from lack of information. Numerical results suggest the effectiveness of our proposed model and solution.

Future research directions include improving the bandit algorithm to select multiple SBSs at every round at lower complexity. In essence, in \cite{Uchiya10:AAB} and \cite{Cesa12:CB}, combinatorial bandit algorithms are proposed that offer lower complexities compared to EXP4;  however, they  cannot be used in case of sleeping arms, where the set of available actions is time-variant. In addition, the model can be adapted to the scenario in which SBSs are selected based on maximum offered rewards (for instance throughput) rather than QoS guarantee.
\bibliographystyle{IEEEbib}
\bibliography{main}

\begin{thebibliography}{10}

\bibitem{Hossain14:ETG}
E.~Hossain, M.~Rasti, H.~Tabassum, and A.~Abdelnasser,
\newblock ``Evolution toward {5G} multi-tier cellular wireless networks: An
  interference management perspective,''
\newblock {\em IEEE Wireless Communications}, vol. 21, no. 3, pp. 118--127,
  June 2014.

\bibitem{Fodor12:DAN}
G.~Fodor, E.~Dahlman, G.~Mildh, S.~Parkvall, N.~Reider, G.~Mikl{\'o}s, and
  Z.~Tur{\'a}nyi,
\newblock ``Design aspects of network assisted device-to-device
  communications,''
\newblock {\em IEEE Communications Magazine}, vol. 50, no. 3, pp. 170--177,
  March 2012.

\bibitem{Zou15:NSDS}
K.J. Zou, K.W. Yang, M.~Wang, B.~Ren, J.~Hu, J.~Zhang, M.~Hua, and X.~You,
\newblock ``Network synchronization for dense small cell networks,''
\newblock {\em IEEE Wireless Communications}, vol. 22, no. 2, pp. 108--117,
  April 2015.

\bibitem{Amin15:APG}
R.~Amin and J.~Martin,
\newblock ``Assessing performance gains through global resource control of
  heterogeneous wireless networks,''
\newblock {\em IEEE Transactions on Mobile Computing}, vol. PP, no. 99, pp.
  1--1, 2015.

\bibitem{Zhang15:ROIM}
H.~Zhang, Y.~Wang, and H.~Ji,
\newblock ``Resource optimization based interference management for hybrid
  self-organized small cell network,''
\newblock {\em IEEE Transactions on Vehicular Technology}, vol. PP, no. 99, pp.
  1--1, 2015.

\bibitem{Zhang15:CIM}
H.~Zhang, C.~Jiang, and J.~Cheng,
\newblock ``Cooperative interference mitigation and handover management for
  heterogeneous cloud small cell networks,''
\newblock {\em IEEE Wireless Communications}, vol. 22, no. 3, pp. 92--99, June
  2015.

\bibitem{Semiari14:MTP}
O.~Semiari, W.~Saad, S.~Valentin, M.~Bennis, and B.~Maham,
\newblock ``Matching theory for priority-based cell association in the downlink
  of wireless small cell networks,''
\newblock in {\em IEEE International Conference on Acoustics, Speech and Signal
  Processing}, May 2014, pp. 444--448.

\bibitem{Namvar14:ACM}
N.~N.~Namvar, W.~Saad, B.~Maham, and S.~Valentin,
\newblock ``A context-aware matching game for user association in wireless
  small cell networks,''
\newblock in {\em IEEE International Conference on Acoustics, Speech and Signal
  Processing}, May 2014, pp. 439--443.

\bibitem{Saad14:CAG}
W.~Saad, Z.~Han, R.~Zheng, M.~Debbah, and H.V. Poor,
\newblock ``A college admissions game for uplink user association in wireless
  small cell networks,''
\newblock in {\em Proceedings IEEE INFOCOM}, April 2014, pp. 1096--1104.

\bibitem{Chen15:JUA}
Y.~Chen, J.~Li, W.~Chen, Z.~Lin, and B.~Vucetic,
\newblock ``Joint user association and resource allocation in the downlink of
  heterogeneous networks,''
\newblock {\em IEEE Transactions on Vehicular Technology}, vol. PP, no. 99, pp.
  1--1, 2015.

\bibitem{Mesodiakaki14:EEU}
A.~Mesodiakaki, F.~Adelantado, L.~Alonso, and C.~Verikoukis,
\newblock ``Energy-efficient user association in cognitive heterogeneous
  networks,''
\newblock {\em IEEE Communications Magazine}, vol. 52, no. 7, pp. 22--29, July
  2014.

\bibitem{Elbassiouny15:TAU}
S.O. Elbassiouny, A.~Elhamy, and A.S. Ibrahim,
\newblock ``Traffic-aware user association technique for dynamic on/off
  switching of small cells,''
\newblock in {\em IEEE Wireless Communications and Networking Conference},
  March 2015, pp. 866--871.

\bibitem{Athanasiou09:ACL}
G.~Athanasiou, T.~Korakis, O.~Ercetin, and L.~Tassiulas,
\newblock ``A cross-layer framework for association control in wireless mesh
  networks,''
\newblock {\em IEEE Transactions on Mobile Computing}, vol. 8, no. 1, pp.
  65--80, Jan 2009.

\bibitem{Ye13:UAL}
Q.~Ye, B.~Rong, Y.~Chen, M.~Al-Shalash, C.~Caramanis, and J.G. Andrews,
\newblock ``User association for load balancing in heterogeneous cellular
  networks,''
\newblock {\em IEEE Transactions on Wireless Communications}, vol. 12, no. 6,
  pp. 2706--2716, June 2013.

\bibitem{Sudeva11:EHS}
S.~Sudevalayam and P.~Kulkarni,
\newblock ``Energy harvesting sensor nodes: Survey and implications,''
\newblock {\em IEEE Communications Surveys Tutorials}, vol. 13, no. 3, pp.
  443--461, March 2011.

\bibitem{Song14:TUA}
Y.~Song, M.~Zhao, W.~Zhou, and H.~Han,
\newblock ``Throughput-optimal user association in energy harvesting
  relay-assisted cellular networks,''
\newblock in {\em International Conference on Wireless Communications and
  Signal Processing}, Oct 2014, pp. 1--6.

\bibitem{Yu15:EHP}
P.-S. Yu, J.~Lee, T.Q.S. Quek, and Y.-W.P. Hong,
\newblock ``Energy harvesting personal cells-traffic offloading and network
  throughput,''
\newblock in {\em IEEE International Conference on Communications}, June 2015,
  pp. 2184--2189.

\bibitem{Lee11:EMS}
P.~Lee, Z.~Ang Eu, M.~Han, and H.~Tan,
\newblock ``Empirical modeling of a solar-powered energy harvesting wireless
  sensor node for time-slotted operation,''
\newblock in {\em IEEE Wireless Communications and Networking Conference},
  March 2011, pp. 179--184.

\bibitem{Sakr15:AMT}
A.~H. Sakr and E.~Hossain,
\newblock ``Analysis of k-tier uplink cellular networks with ambient {RF}
  energy harvesting,''
\newblock {\em IEEE Journal on Selected Areas in Communications}, vol. PP, no.
  99, pp. 1--1, 2015.

\bibitem{Amari97:CFE}
S.V. Amari and R.B. Misra,
\newblock ``Closed-form expressions for distribution of sum of exponential
  random variables,''
\newblock {\em IEEE Transactions on Reliability}, vol. 46, no. 4, pp. 519--522,
  1997.

\bibitem{Yao90:OPA}
Y.D. Yao and A.U.H. Sheikh,
\newblock ``Outage probability analysis for microcell mobile radio systems with
  cochannel interferers in {R}ician/{R}ayleigh fading environment,''
\newblock {\em Electronics Letters}, vol. 26, no. 13, pp. 864--866, 1990.

\bibitem{Billingsley86:PaM}
P.~Billingsley,
\newblock {\em Probability and Measures},
\newblock {W}iley, 2nd edition, 1986.

\bibitem{Papoulis03}
A.~Papoulis and S.U. Pillai,
\newblock {\em Probability, Random Variables, and Stochastic Processes},
\newblock {T}ata {M}c{G}raw-{H}ill, 1st edition, 2002.

\bibitem{Kleinberg10:RBS}
R.~Kleinberg, A.~Niculescu-Mizil, and Y.~Sharma,
\newblock ``Regret bounds for sleeping experts and bandits,''
\newblock {\em Machine Learning}, vol. 80, no. 2-3, pp. 245--272, 2010.

\bibitem{Auer03:NMB}
P.~Auer, N.~Cesa-Bianchi, Y.~Freund, and R.E. Schapire,
\newblock ``The nonstochastic multiarmed bandit problem,''
\newblock {\em SIAM Journal on Computing}, vol. 32, no. 1, pp. 48--77, Jan.
  2003.

\bibitem{Bianchi06:PLG}
N.~Cesa-Bianchi and G.~Lugosi,
\newblock {\em Prediction, Learning, and Games},
\newblock Cambridge University Press, 2006.

\bibitem{Lin15:OUA}
Y.~Lin, W.~Bao, W.~Yu, and B.~Liang,
\newblock ``Optimizing user association and spectrum allocation in hetnets: A
  utility perspective,''
\newblock {\em IEEE Journal on Selected Areas in Communications}, vol. 33, no.
  6, pp. 1025--1039, June 2015.

\bibitem{Son11:BSO}
K.~Son, H.~Kim, Y.~Yi, and B.~Krishnamachari,
\newblock ``Base station operation and user association mechanisms for
  energy-delay tradeoffs in green cellular networks,''
\newblock {\em IEEE Journal on Selected Areas in Communications}, vol. 29, no.
  8, pp. 1525--1536, September 2011.

\bibitem{Uchiya10:AAB}
T.~Uchiya, A.~Nakamura, and M.~Kudo,
\newblock ``Algorithms for adversarial bandit problems with multiple plays,''
\newblock in {\em Algorithmic Learning Theory}, Oct 2010, pp. 375--389.

\bibitem{Cesa12:CB}
N.~Cesa-Bianchia and G.~Lugosi,
\newblock ``Combinatorial bandits,''
\newblock {\em Journal of Computer and System Sciences}, vol. 78, no. 5, pp.
  1404–1422, 2012.

\end{thebibliography}
\end{document}